\documentclass[a4paper,11pt]{article}
\usepackage{pos}
\usepackage{graphicx}

\title{On the potential of widening the band of X-ray polarimetry}

\author*[a]{Nicolas De Angelis}
\affiliation[a]{INAF-IAPS,\\
  100 via del Fosso del Cavaliere, Rome, Italy}
\emailAdd{nicolas.deangelis@inaf.it}

\abstract{The Imaging X-ray Polarimetry Explorer (IXPE) has made X-ray polarimetry a routine astrophysical diagnostic, but its sensitivity is confined to the 2--8~keV band, where the photoelectric effect in a gas is most efficiently exploited. Several of the questions raised by its results, as well as some classical goals of high-energy astrophysics, require polarimetry below 2~keV, to reach the thermal emission of neutron stars, black-hole discs and blazars, and above 8~keV, where reflection, cyclotron lines and the non-thermal emission of supernova remnants, pulsar wind nebulae and magnetars can be isolated. Following a recent review of the subject, we discuss the scientific case for extending the band over roughly 0.1--80~keV and beyond, and the techniques that can deliver it: Bragg diffraction on graded multilayers, photoelectric gas imagers read out by pixel ASICs with time-of-arrival capability, time projection chambers and focal-plane Compton polarimeters, including their combination in stacked focal planes. We then summarise how these techniques are being implemented, or could be, in REDSoX and GOSoX, eXTP, XPP, WXPT and in EXPO and PHEMTO, proposed to the ESA M8 call, as well as in the wide-field transient polarimeter POLAR-2, and we discuss the trade-offs that will drive the design of a wide-band polarimetry mission.}

\FullConference{Multifrequency Behaviour of High Energy Cosmic Sources - XV (MULTIF2025)\\
9--14 June, 2025\\
Mondello, Palermo, Italy\\}

\begin{document}
\maketitle

\section{Scientific motivation for high-energy wide-band polarimetry}
\label{sec:science}

For almost five decades after the first detection of the polarised emission of the Crab Nebula in the 1970s \cite{novick1972,weisskopf1978}, X-ray polarimetry remained essentially unexplored. The Imaging X-ray Polarimetry Explorer (IXPE), a NASA/ASI Small Explorer launched on 9 December 2021 \cite{weisskopf2022,soffitta2021instrument}, has changed this: each of its three telescopes couples a grazing-incidence mirror to a Gas Pixel Detector (GPD) \cite{baldini2021}, an imaging photoelectric polarimeter sensitive in the 2--8~keV band. This band was chosen because celestial sources are bright there, the optics are effective, and most source classes were expected to be polarised to some degree \cite{soffitta2024galaxies}. IXPE has observed about one hundred sources, and its operations have been extended through 2028 \cite{soffitta2026lessons}. It has measured the polarisation of pulsar wind nebulae (PWNe) \cite{bucciantini2023}, supernova remnants (SNRs) \cite{vink2022}, magnetars \cite{taverna2022}, accreting pulsars \cite{doroshenko2022}, blazars \cite{liodakis2022,digesu2022}, Seyfert galaxies \cite{gianolli2023,ursini2023} and the molecular clouds of the Galactic Centre \cite{marin2023}, often at odds with the expectations.

These results establish polarimetry as a diagnostic complementary to imaging, spectroscopy and timing, but they also show the limits of a single narrow band, which is set by the energy dependence of the photoelectric cross-section and of the photoelectron range in the gas \cite{soffitta2024galaxies}. Following the review of Ref.~\cite{soffitta2024galaxies} and updating it with the developments of the last two years, we discuss the science that lies outside the IXPE band (this section), the techniques that can reach it (Sect.~\ref{sec:tech}) and the projects that are putting them to use (Sect.~\ref{sec:outlook}).

\subsection{The soft band (below 2~keV)}

Isolated neutron stars have thermal spectra peaking at 0.1--0.3~keV, where their flux is orders of magnitude larger than in the IXPE band: about 100 of the $\sim$1800 known rotation-powered pulsars are thermal X-ray emitters, with temperatures of $10^5$--$10^6$~K. Their strong magnetic fields are expected to polarise the radiation \cite{pavlov2000}, which would give access to the geometry of the emitting regions \cite{soffitta2024galaxies}. The same holds for magnetars, about ten of which lie at high Galactic latitude \cite{soffitta2024galaxies}. IXPE has so far observed seven magnetars, revealing complex emission regions and, often, a strong energy dependence of the polarisation \cite{taverna2022,zane2023,turolla2023,heyl2024,rigoselli2025,stewart2025,stewart2026,taverna2026,taverna2026ep}. The QED effect of vacuum birefringence is still debated: the high polarisation of the radio magnetar 1E~1547.0$-$5408 has been interpreted as its signature \cite{stewart2026}, while an independent analysis concludes that it is not compelling evidence on its own \cite{taverna2026}. A larger sample, and the band where the thermal emission peaks, would permit a cleaner test.

Among black-hole binaries, about a dozen lie at high Galactic latitude. In the high-soft state the disc emits thermally, and polarimetry below 1~keV constrains its magnetic field through Faraday rotation and depolarisation \cite{davis2009} and may reveal warped discs \cite{soffitta2024galaxies}. High-synchrotron-peaked (HSP) blazars, which also lie at high latitude, peak between 0.1 and 10~keV; IXPE found X-ray polarisation degrees two to three times larger than at longer wavelengths \cite{liodakis2022,digesu2022}, and, in Mrk~421, a rapid rotation of the polarisation angle that was not accompanied by a rotation at longer wavelengths \cite{digesu2023}. A measurement in the soft band would test whether the transition between these behaviours is smooth. Because of interstellar absorption the soft band is useful mainly at high Galactic latitude, which is the case for all the classes above.

\subsection{The classical band (2--8~keV): sensitivity, resolution and agility}

Even in the IXPE band, the sample of most classes of sources remains limited to the brightest objects \cite{soffitta2024galaxies}. An effective area about ten times larger would extend the sample to more Seyfert-1 galaxies, whose primary emission is due to Comptonisation in a hot corona \cite{gianolli2023}, and to low-synchrotron-peaked (LSP) blazars, for which IXPE has obtained only upper limits and a secure detection would clarify whether leptons or hadrons are accelerated \cite{zhang2013}. The angular resolution of IXPE, 25--30~arcsec, was fully exploited only for the Crab PWN \cite{bucciantini2023}. Optics with 5--10~arcsec and a few thousand cm$^2$ are becoming accessible, thanks to silicon pore optics and replica techniques, and an improvement by a factor 3--6 would allow detailed studies of the morphology of PWNe, bringing at least seven more within reach, and would probe turbulence and field orientation closer to the shocks of SNRs \cite{soffitta2024galaxies}. A gas detector with an energy resolution twice better than IXPE appears feasible, and would allow the thermal lines of SNRs to be excluded and the Fe K$\alpha$ line, continuum and edge of AGN to be separated \cite{soffitta2024galaxies}. Finally, IXPE, a Small Explorer, needs days to re-point, which prevents polarimetry of fast transients and of the early phases of bursts; EXPO, proposed to ESA, addresses this with a wide-field monitor that triggers an autonomous re-pointing within about 50~s \cite{soffitta2026expo}.

\subsection{The hard band (above 8~keV)}

The IXPE band borders several of the main objectives of polarimetry \cite{krawczynski2011}. Above 6~keV the spectra of Seyfert-1 galaxies are dominated by reflection from the disc, and those of Seyfert-2 galaxies by scattering from the ionisation cones or from the torus. In NGC~4151 the reflection fraction is only 8\% in the IXPE band and doubles at higher energies, so the polarisation of the reflected component, which would give the geometry of the disc and possibly the spin of the black hole, has not been firmly measured \cite{gianolli2023}. In the reflection-dominated Circinus galaxy the reflection from the torus was measured, but the polarisation from the ionisation cone was largely unconstrained \cite{ursini2023}, and the same holds for NGC~1068 \cite{marin2024ngc1068}; a possible misalignment between the two structures could be measured with efficient spectro-polarimetry over a wide band \cite{goosmann2011}. For SNRs, lines and thermal emission hamper the 2--8~keV band \cite{vink2022}, while imaging polarimetry above the iron lines would isolate the synchrotron emission and may explain the features detected by NuSTAR in Cassiopeia~A \cite{grefenstette2015}. Magnetars have a hard tail extending at least to 100~keV \cite{gotz2006,makishima2021}, and polarimetry at 6--30~keV would explore its transition from the soft component. As the tail is faint, the background must be kept low, and imaging does so by limiting it to the small area of the point-spread function of the optics. Accreting pulsars are less polarised than expected \cite{doroshenko2022}, and the polarisation at the cyclotron line, which is mostly above 10~keV, is still unmeasured: it is not clear whether it is large ($\sim$80\%) \cite{meszaros1988} or zero \cite{caiazzo2021a,caiazzo2021b}. Finally, a hard X-ray imager with a resolution of 10--20~arcsec would make the tomography of the Galactic Centre clouds \cite{marin2023} cleaner and extend it to more clouds.

The first measurements above the IXPE band come from XL-Calibur, a balloon-borne focal-plane Compton polarimeter with a 15--80~keV bandpass \cite{abarr2021} that flew from Esrange to Canada in July 2024. It measured a polarisation degree of $(25.1\pm2.9)\%$ at an angle of $(129.8\pm3.2)^\circ$ for the Crab in 19--64~keV \cite{awaki2025crab}, and $5.0^{+2.7}_{-3.0}\%$ for Cyg~X-1 in the hard state, at an angle aligned with the radio jet \cite{awaki2025cyg}; the latter is a hint rather than a detection (a chance probability of 8.7\% after 43~hours). A balloon can reach only the brightest sources, and a satellite is needed to bring hard X-ray polarimetry to the general population. Cyg~X-1 is a case in point: INTEGRAL found its emission above 400~keV to be strongly polarised and probably related to the jet \cite{laurent2011}, while the polarisation degree measured by IXPE in 2--8~keV is only $4.0\pm0.2$\% \cite{krawczynski2022}; polarimetry in the intermediate range is needed to connect the two regimes.

\section{Available or in-development technologies}
\label{sec:tech}

\begin{figure}[t]
  \centering
  \includegraphics[width=\textwidth]{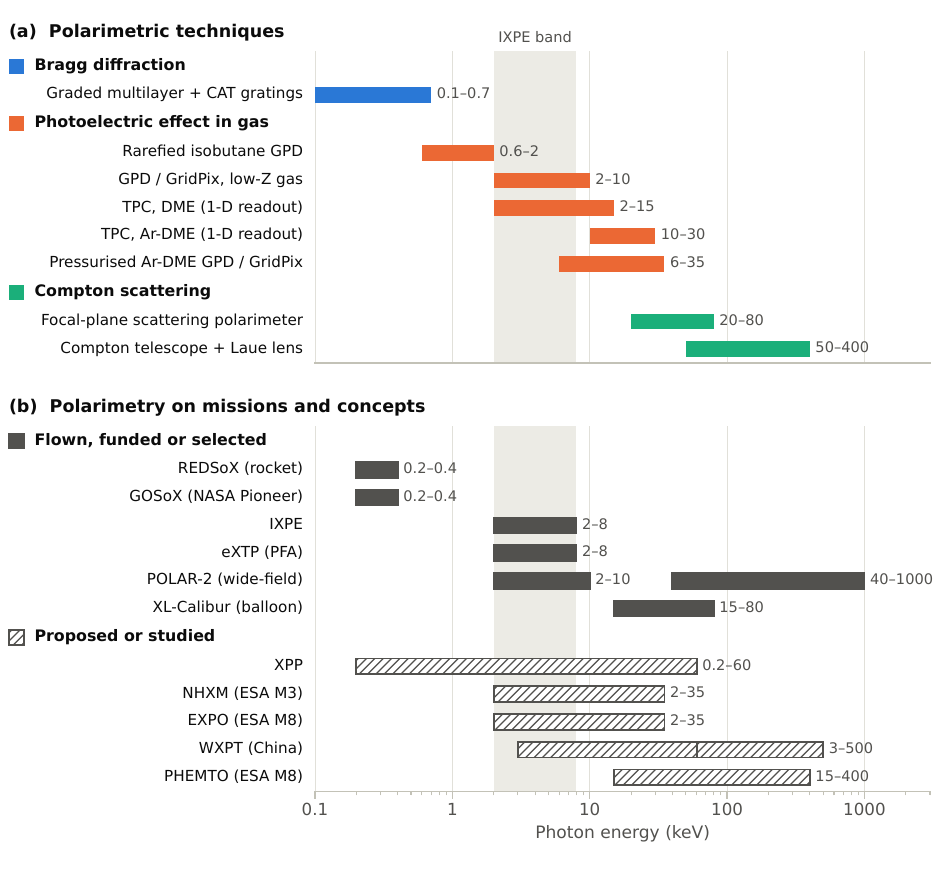}
  \caption{Energy coverage of (a) the polarimetric techniques of Sect.~\ref{sec:tech} and (b) the missions and concepts discussed in the text. The shaded band is the 2--8~keV range of IXPE. Ranges are indicative and are taken from the literature cited in the text. Solid bars in (b) denote missions that have flown, are funded or have been selected; hatched bars denote concepts that are proposed or under study. The two bars of POLAR-2, a wide-field instrument without optics, is drawn with its low- and high-energy polarimeters (LPD and HPD); the WXPT bar is divided at 60~keV, the upper limit of its focal-plane polarimeters, above which a wide-field polarimeter extends the band to 500~keV.}
  \label{fig:bands}
\end{figure}

No single technique covers the range from 0.1 to 80~keV optimally, because the best-suited process changes with energy: Bragg diffraction below $\sim$0.7~keV, the photoelectric effect in a gas from $\sim$0.6 to $\sim$35~keV, and Compton scattering above $\sim$20~keV (Fig.~\ref{fig:bands}a). The sensitivity is conveniently expressed by the minimum detectable polarisation at 99\% confidence \cite{weisskopf2010},
\begin{equation}
  \mathrm{MDP}_{99} = \frac{4.29}{\mu\,R_S}\sqrt{\frac{R_S+R_B}{T}},
  \label{eq:mdp}
\end{equation}
where $\mu$ is the modulation factor, $R_S=F\,A\,\varepsilon$ the source rate ($F$ photon flux, $A$ effective area of the optics, $\varepsilon$ detection efficiency), $R_B$ the background rate in the source region and $T$ the exposure. For a negligible background $\mathrm{MDP}_{99}=4.29/(\mu\sqrt{F A\varepsilon T})$, so that the quality factor $\mu\sqrt{\varepsilon}$ sets the sensitivity together with the area of the optics. In a gas detector the two factors are traded against each other, because a longer drift increases the efficiency but also the diffusion, which blurs the tracks \cite{soffitta2024galaxies}. Figure~\ref{fig:bands}b shows the bands of the missions and concepts that use these techniques (Sect.~\ref{sec:outlook}).

\subsection{Bragg diffraction on graded multilayers (0.1--0.7~keV)}

At a Bragg angle close to 45$^\circ$ the component of the electric field in the plane of incidence is almost entirely suppressed, so that the reflected intensity is modulated with an amplitude close to 100\% as the reflector rotates about the beam. Natural crystals were limited by their narrow bandwidth and by the few that diffract at 45$^\circ$ in 2--8~keV; artificial multilayers made soft X-rays accessible, and a graded thickness widened the bandwidth and hence the efficiency \cite{soffitta2024galaxies}. The most advanced design \cite{marshall2017} combines laterally graded multilayer mirrors (LGML) with critical-angle transmission (CAT) gratings placed close to an X-ray mirror: the grating disperses the radiation by wavelength, the grading of the multilayer matches the dispersion, and the radiation reflected at nearly 45$^\circ$ is recorded by a CCD along the grading direction. The advantages are a wide soft band (0.1--0.7~keV), a modulation approaching 100\% throughout, a very good energy resolution thanks to the dispersion, a low background, since the signal amplitude can be correlated with its position on the CCD, and a geometry that permits stacking (Sect.~\ref{sec:stack}) with little loss of mirror area at higher energies \cite{soffitta2024galaxies}. A single reflection measures one linear-polarisation component, so that three devices at different orientations are needed to measure $I$, $Q$ and $U$ simultaneously, or one device must rotate, at the expense of the sensitivity to variable sources.

The concept is being demonstrated by REDSoX, a NASA-funded sounding rocket (five replicated Ni Wolter-I shells of 2.5~m focal length, 48 CAT gratings of 200~nm period, three LGMLs at 120$^\circ$ to each other) whose first flight, on Mrk~421, is planned for late 2027 \cite{heine2026}. Its orbital version, GOSoX, has been selected as a NASA Astrophysics Pioneer for launch in 2030 \cite{marshall2026}: a SmallSat for the 0.2--0.4~keV band with $E/\Delta E\sim100$ and an expected MDP of 3--10\% on more than a dozen targets in one year, rolling at no more than 1$^\circ$/s.

\subsection{Rarefied gas below 2~keV (0.6--2~keV)}

Below 2~keV the photoelectron range shortens rapidly. A rarefied mixture based on, e.g., isobutane can provide some imaging sensitivity below 1~keV, and was studied for the XEUS mission \cite{pacciani2003}. The diffusion grows as $\sqrt{d_l/P}$ when the pressure $P$ decreases, so the drift gap $d_l$ must be short (0.5~cm or less), with the possible drawback of a larger fraction of events with incomplete charge collection; helium diffuses too much and would need a quencher; and polarisation-insensitive Auger electrons set a lower limit of about 600~eV, from the K-shell energy of carbon. Such a device would be virtually transparent above 2~keV, and a thin silicon-nitride window would perform better than polypropylene or polyimide ones \cite{soffitta2024galaxies}.

\subsection{Gas pixel detectors and GridPix (2--10~keV)}
\label{sec:gpd}

The Gas Pixel Detector \cite{costa2001,bellazzini2007} opened polarimetry in the classical band: the track of a photoelectron, drifted to a multiplication stage and read out by a pixel ASIC, gives the impact point and the emission angle, which is peaked along the polarisation direction. The IXPE detectors \cite{baldini2021} have a 1-cm drift gap with 800~mbar of dimethyl ether (DME), a gas electron multiplier (GEM) and a CMOS ASIC of 105\,600 hexagonal pixels at 50~$\mu$m pitch, an optimum found before the selection of IXPE that a fine tuning is not expected to change \cite{soffitta2024galaxies}. The main shortcoming is the dead time of 1.1~ms, which the XPOL-III ASIC reduces to 0.15~ms for eXTP \cite{minuti2023}; with optics of $\sim$1500~cm$^2$, like XMM-Newton, or larger, a more substantial improvement is needed \cite{soffitta2024galaxies}.

The ASICs of the Medipix family, Timepix3 \cite{poikela2014} and Timepix4 \cite{llopart2022}, record in each pixel the time of arrival (ToA, with a nominal resolution of 1.5625~ns and 195~ps, respectively) and the time over threshold (ToT, proportional to the energy). Timepix3 has $256\times256$ square pixels of $55~\mu$m and is virtually dead-time free up to 40\,Mhits\,cm$^{-2}$\,s$^{-1}$, while Timepix4 could extend the field of view or preserve it at a longer focal length. The ToA yields three-dimensional tracks, which should improve the photoelectron direction and thus the modulation factor. In GridPix \cite{vandergraaf2007,ligtenberg2018}, a metallic mesh (InGrid) on dielectric pillars 50~$\mu$m high provides the multiplication without additional diffusion in the transfer gap (8 against 55~$\mu$m rms for the GEM; Table~\ref{tab:gpd}), and removes the pixel-dependent, rate-dependent loss of gain caused by charge attachment on the GEM, which is hard to calibrate in flight. Three issues remain open \cite{soffitta2024galaxies}. (i) The absorption of DME by the detector body, probably by the glue, caused a $\sim$20\% loss of efficiency; it may be solved with careful materials or with another low-diffusion, low-$Z$ mixture (isobutane, CO$_2$). (ii) The mesh requires a thin Si$_3$N$_4$ coating of the ASIC as a protection against sparks, whose effect on charging has to be evaluated. (iii) It must be shown that square, slightly larger pixels do not add a spurious modulation, or that it can be modelled and subtracted, as was done for the GPD with an extensive calibration campaign.

\begin{table}[t]
  \centering
  \begin{tabular}{lcc}
    \hline
     & IXPE GPD (XPOL-III) & GridPix (Timepix3) \\
    \hline
    Pixel pattern & hexagonal, 50~$\mu$m pitch & square, $55\times55~\mu$m$^2$ \\
    Signal output & analog & digital (ToT) \\
    Track imaging & 2-D & 3-D \\
    Diffusion in transfer gap (DME, $\mu$m rms) & 55 & 8 \\
    Noise per pixel ($e^-$ rms) & 50 (30) & 60 \\
    Minimum threshold ($e^-$) & 2300 (150) & 500 (ToT/ToA) \\
    \hline
  \end{tabular}
  \caption{IXPE GPD and GridPix with a Timepix3 ASIC, adapted from Ref.~\cite{soffitta2024galaxies}. Values in parentheses are those expected for XPOL-III.}
  \label{tab:gpd}
\end{table}

\subsection{Time projection chambers (2--30~keV)}

Black et al.\ devised a detector that decouples the modulation factor from the quantum efficiency, the time projection chamber (TPC) \cite{black2007}. The drift field is perpendicular to the optical axis and a one-dimensional readout sits on a side, out of the absorption path, so that the drift length, and hence the diffusion, is constant along the depth of the cell and a very thick, efficient polarimeter is possible. The drift time provides the second coordinate of the track, but the start time is unknown, so the interaction point is not localised, and the device must rotate around the beam to cancel its systematic effects. The drawbacks are the lack of imaging, which excludes the extended sources that IXPE has shown to be essential (SNRs \cite{vink2022}, PWNe \cite{bucciantini2023}, molecular clouds \cite{marin2023}); a higher background, which cannot be subtracted from the same exposure and is blurred by the absorption of a converging beam in a thick detector \cite{dimarco2023}; and the need for rotation, which complicates the study of sources with erratic periodicity and calls for multiple devices \cite{soffitta2024galaxies}. An adequate level of readiness was reached, but building the flight device for GEMS \cite{hill2012} proved harder than expected and the mission was cancelled for programmatic reasons. A TPC may still be considered for a stacked configuration (Sect.~\ref{sec:stack}), its sensitive plane being parallel to the photon beam.

\subsection{Pressurised argon imagers (6--35~keV)}

Sensitivity above 6~keV, up to 30--35~keV, requires pressurised argon and a thick cell. A GPD with a 2-bar, 2-cm Ar-DME mixture has been tested \cite{fabiani2012}, and a 3-cm, 3-bar implementation was proposed for the New Hard X-ray Mission (NHXM) of the ESA M3 call \cite{soffitta2010,tagliaferri2012}, which had four multilayer telescopes of 10~m focal length, an IXPE-like low-energy polarimeter (LEP, 2--10~keV) and this medium-energy one (MEP, 5--35~keV). With similar optics areas ($\sim$500 against $\sim$600~cm$^2$ for IXPE), the computed MDP of the LEP for a 0.5~mCrab source in 10~days was 5.1\% against 5.5\% for IXPE, and the MEP was $\sim$34\% worse in mCrab, so that one LEP and two MEPs would be well balanced \cite{soffitta2024galaxies}. A 7.8-cm thick TPC with 1~bar of Ar-DME (60:40) was also tested up to 16~keV \cite{takeda2020}. At high energies the instrumental background, mainly due to the Earth albedo, decays with energy less steeply than celestial spectra, so that the rejection must improve: through the topology of the tracks \cite{dimarco2023}, as with proportional counters, and possibly with an active anticoincidence for minimum-ionising particles, all of which is viable with GridPix. An imaging detector also measures the background in the same sky image, within the point-spread function of the optics, orders of magnitude smaller than the area of a non-imaging device \cite{soffitta2024galaxies}.

\subsection{Focal-plane Compton polarimeters (20--80~keV and beyond)}
\label{sec:compton}

Above $\sim$20~keV photoelectric polarimetry with argon mixtures starts to compete with Compton polarimetry, with a possible overlap at 20--30~keV; krypton mixtures, still to be studied, may be discouraged by their large fluorescence yield. Focal-plane Compton polarimeters have so far flown only on balloons, X-Calibur \cite{abarr2020} and XL-Calibur \cite{abarr2021}, and were proposed for PolariS \cite{hayashida2016} and for the NASA SMEX PolSTAR \cite{krawczynski2016}. They have two stages \cite{delmonte2023}: a scatterer of low atomic number (lithium, beryllium or plastic scintillator \cite{costa2024}; organic crystals proved less effective than expected \cite{fabiani2013}) and an absorber. A passive configuration can have a threshold of 5--6~keV on a satellite, or 15--20~keV on a balloon \cite{soffitta2024galaxies}, but its background is high and needs an efficient active shield; XL-Calibur has a beryllium scatterer surrounded by four planes of CZT detectors and a thick BGO shield \cite{iyer2023}. An active configuration has a much smaller background from coincidences alone, but a higher threshold, $\sim$20~keV even on a satellite \cite{fabiani2013}. A single scatterer does not image; with segmented scatterers (16--36 elements of $\sim$4--9~mm$^2$) at the focus of $\sim$10-m multilayer optics a coarse imaging of $\sim$1~arcmin is possible, which is the configuration proposed for PolariS. Reading both sides of scatterer and absorber with thinned silicon drift detectors or SiPMs \cite{muleri2024} could lead to a three-dimensional Compton polarimeter with a lower background and a higher modulation. The sensitivity is hard to predict: a study for NHXM \cite{soffitta2010}, with an active plastic scatterer, a cylindrical absorber, an assumed triggering efficiency and no background, gave an MDP of 6\% for a 1.8~mCrab source, a figure to be validated by simulations, tests and the XL-Calibur data.

Beyond the $\sim$80~keV reached by multilayer mirrors, focusing relies on Laue lenses, which diffract in transmission through crystals (bent Si and Ge, for PHEMTO). PHEMTO combines such a lens with a multilayer mirror in front of a stacked Si/CdTe Compton telescope, which measures the polarisation on double Compton events \cite{laurent2021} (Sect.~\ref{sec:outlook}).

\subsection{Stacked configurations}
\label{sec:stack}

A mission spanning 0.1--80~keV can be a multi-instrument payload, with with a telescope (mirror and detector) optimised for each band, or a single telescope that exchanges the instrument in the focal plane. The first optimises each telescope but divides the area; the second saves mirrors but covers the band less efficiently and not simultaneously \cite{soffitta2024galaxies}. A third option, given high-throughput, fine-resolution multilayer mirrors, is to stack detectors of complementary sensitivity along the optical axis, so that the photons not absorbed by one reach the next. This has been proposed \cite{jahoda2019,soffitta2021ngxp}, and a Compton polarimeter behind the LEP was already part of NHXM \cite{soffitta2010}.

\begin{figure}[t]
  \centering
  \includegraphics[width=\textwidth]{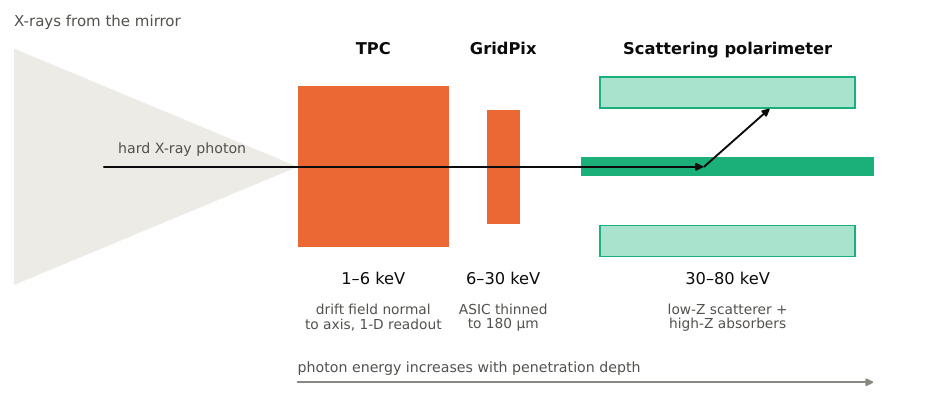}
  \caption{Sketch of the three-stage stacked focal-plane polarimeter proposed in Ref.~\cite{soffitta2024galaxies}: a TPC tuned to 1--6~keV, with the drift field normal to the optical axis and a one-dimensional readout, followed by a thinned GridPix imager for 6--30~keV and by a scattering polarimeter, a low-$Z$ scatterer surrounded by high-$Z$ absorbers, for 30--80~keV. Not to scale.}
  \label{fig:stack}
\end{figure}

A natural binary system is a TPC on top (DME for 2--15~keV, or Ar-DME for 10--30~keV) coupled through a thin Be window to a scattering polarimeter that extends the band to 80~keV, natural because the sensitive plane of the TPC is parallel to the beam; without imaging, however, it loses a significant part of the science of Sect.~\ref{sec:science} \cite{soffitta2024galaxies}. An imaging stack needs a sensitive plane perpendicular to the beam and detectors thin enough to limit the absorption in passive materials, namely the ASIC and its board. The Timepix3 wafer, originally 800~$\mu$m thick, can be thinned to 180~$\mu$m without loss of performance, which gives a transparency of 85\% at 20~keV \cite{soffitta2024galaxies}. With a thin board, or a hole beneath the ASIC, and possibly a Be window, a stacked imaging system is feasible: the gas stage must be at the focus, while the scattering stage tolerates some defocusing, given the small convergence of hard X-ray optics. The more ambitious three-stage stack of Fig.~\ref{fig:stack} puts a TPC, a GridPix and a scattering stage on a single telescope. A design in which low-energy (2--8~keV) or medium-energy (6--30~keV) imaging photoelectric polarimeters are stacked with an active Compton polarimeter (20--80~keV), alongside a spectro-imaging camera covering 0.3--80~keV, has recently been described in detail \cite{muleri2026}.

\section{Conclusions and Outlook}
\label{sec:outlook}

The science cases of Sect.~\ref{sec:science} and the techniques of Sect.~\ref{sec:tech} are converging in a set of projects which, taken together, cover the range from 0.2~keV to several hundred keV (Fig.~\ref{fig:bands}b). We summarise them according to their status, and then discuss the trade-offs that they illustrate.

\subsection{Projects}

\paragraph{Flown, funded or selected.} IXPE \cite{weisskopf2022}, whose operations are extended through 2028 \cite{soffitta2026lessons}, will remain for some years the only focusing polarimeter in the classical band. The Chinese-led eXTP, planned for launch in 2030 \cite{zhang2019,zhou2025}, will carry a Polarimetry Focusing Array of GPDs read out by the XPOL-III ASIC, with a larger throughput but a similar energy range \cite{soffitta2026expo}. In the soft band, REDSoX is scheduled to fly in late 2027 and GOSoX is planned for launch in 2030 (Sect.~\ref{sec:tech}), so that a first extension to 0.2--0.4~keV will be available in the same years. At high energies XL-Calibur has provided the first focal-plane measurements above 15~keV (Sect.~\ref{sec:science}).

\paragraph{Proposed concepts.} The X-ray Polarization Probe (XPP) is a NASA probe-class concept with polarimetry from 0.2 to 60~keV \cite{jahoda2019}. The Wide-band X-ray Polarization Telescope (WXPT) is a small (about 500~kg) Chinese concept, proposed in 2017 by the Institute of High Energy Physics for launch in the 2030s, in which multilayer mirrors with a focal length of 10.16~m focus X-rays successively on a TPC polarimeter (3--10~keV) and a Compton polarimeter (10--60~keV), while a wide-field soft gamma-ray polarimeter extends the band to 500~keV \cite{yin2022}. The mirrors are based on silicon pore optics with depth-graded W/Si multilayers, which reach a reflectivity of about 25\% at 20--30~keV \cite{cai2026}.

Two concepts have been proposed to the M8 call of ESA. The Enhanced X-ray Polarimetry Observatory (EXPO) \cite{soffitta2026expo}, proposed for launch around 2041, is built around three pillars: polarimetry over 2--35~keV, a Swift-like autonomous re-pointing within about 50~s, and simultaneous spectroscopy. Its polarimeters are GridPix detectors (Timepix3 with an InGrid, Sect.~\ref{sec:gpd}) of two types: two low-energy polarimeters, with a 1-cm cell of Ne/DME (80/20) at 1~bar, for 2--10~keV, and three medium-energy polarimeters, with a 3-cm cell of Ar/DME (60/40) at 3~bar and a GAGG anticoincidence, for 6--35~keV. They sit behind identical nested Wolter-I mirror modules of 7.5~m focal length, with a half-energy width of 10~arcsec at 1~keV that grows to 14~arcsec at 10~keV and an effective area about three times that of IXPE in the overlapping band. A sixth telescope hosts a spectral imaging camera, a CMOS detector stacked on a CZT one, for simultaneous spectroscopy up to 35~keV, and a wide-field coded-mask instrument derived from SVOM/ECLAIRs (4--150~keV, about 2~sr) triggers the autonomous re-pointing. The Polarimetric High Energy Modular Telescope Observatory (PHEMTO) \cite{laurent2021} is a hard X-ray spectro-imaging polarimeter. In its M8 version a mirror coated with Pt/C and W/Si multilayers (1--80~keV) and a Laue lens (50--400~keV) share a focal length of 20~m, obtained with an extendable boom, and focus on a stacked Si/CdTe Compton telescope, in which polarimetry is obtained from double Compton events above $\sim$15~keV. Its aims include the synchrotron cut-offs of young SNRs, the hard tails of magnetars, and bridging the spectral gap between IXPE and INTEGRAL. Both concepts were shortlisted, among ten proposals, for the second step of the evaluation, but neither was among the four candidates that ESA retained in November 2026 for initial studies, ahead of the M8 selection expected in 2030 \cite{esa2026}.

The two concepts are complementary: EXPO builds on the photoelectric technique and on IXPE's heritage to widen the band at the lower end of the hard X-rays and to add speed, while PHEMTO moves to the Compton regime, with focusing up to 400~keV. Stacking a Compton stage behind the gas imagers, as in the designs of Sect.~\ref{sec:stack}, would bring the band of a payload of the EXPO type to $\sim$80~keV.

\paragraph{Wide-field transient polarimeters.} Widening the band is also considered for non-focusing, wide-field polarimeters devoted to transients. The prompt emission of gamma-ray bursts (GRBs) peaks at hundreds of keV to MeV, but its polarisation is expected to depend on energy, and differently in the competing models: in the dissipative photosphere model it rises towards low energies, where synchrotron photons dominate, and vanishes at the MeV peak and above \cite{lundman2018}, whereas in synchrotron models, where the maximum polarisation degree grows as the local spectrum steepens \cite{gill2020}, it can instead increase towards high energies \cite{lan2021}. Testing this requires polarimetry across the X-ray and gamma-ray ranges of the same burst, which the Compton polarimeter POLAR, with little sensitivity at tens of keV, could not provide \cite{kole2026}. POLAR-2 \cite{deangelis2023, kole2026}, planned for launch around 2028 on the China Space Station, therefore combines a Compton polarimeter (HPD; plastic scintillators read out by silicon photomultipliers, 40--1000~keV) with a gas-based photoelectric polarimeter (LPD; 2--10~keV, field of view of about $90^\circ\times90^\circ$), the same two processes and the same division of roles as in Sect.~\ref{sec:tech}. The charge-dissipating gas microchannel plate of the LPD copes with the high and rapidly varying fluxes of transients, and will provide the first wide-field survey of soft X-ray polarisation. Lacking optics, the sensitivity is limited to bright bursts: about ten GRBs per year are expected with an LPD MDP below 100\%, and one to two per year jointly measured by both polarimeters with the sensitivity needed to test the model \cite{kole2026}. This wide-field route complements the fast re-pointing of a large focusing telescope, as proposed for EXPO.

\paragraph{Other bands.} At gamma-ray energies, where pair production dominates, the polarisation is encoded in the azimuthal distribution of the electron-positron pair. The TPC technique of Sect.~\ref{sec:tech} is used there by HARPO, a TPC measured as a gamma-ray telescope and polarimeter \cite{gros2018}, and emulsion telescopes, which record the pairs with a very high precision, are being developed for sub-GeV/GeV gamma rays by the GRAINE project \cite{nakamura2024}.

\subsection{Discussion}

The projects above embody different answers to the same questions, and the choices that they make are, in our view, the real issue for the design of a wide-band polarimetry mission.

\paragraph{Dedicated telescopes and stacked focal planes.} The two options are not exclusive, and EXPO uses both. Its six mirror modules are identical and what is tuned to a band is the detector: the gas mixture, pressure and depth that are optimal for 2--10~keV are not those optimal for 6--35~keV, so that two telescopes are devoted to the first band and three to the second. Each polarimeter is thus optimised, at the price of a smaller mirror area per band, while the spectral imaging camera stacks a CMOS and a CZT detector in the same focus, so that the photons not absorbed by the first reach the second and a single mirror serves 0.5--35~keV. Stacking the polarimeters themselves keeps the full area of the mirrors for every band, at the price of the transparency of the front stages and of a more demanding detector design (thinned ASICs, windows, defocusing tolerance of the rear stage; Sect.~\ref{sec:stack}). This was the choice of the first EXPO design, in which a Compton polarimeter was stacked behind each gas detector \cite{muleri2026}. Imaging is the dividing line: a TPC reaches high efficiency and modulation, but without imaging it forgoes most of the extended-source science of Sect.~\ref{sec:science} and needs rotation for smoothing out spurious modulation effects, while a GridPix stack preserves the imaging, and with it the background subtraction within the point-spread function of the optics.

\paragraph{Overlap and calibration.} The lesson of IXPE is that systematic effects, spurious modulation first, are controlled by careful calibration, on the ground and in flight with polarised and unpolarised sources \cite{soffitta2026lessons}. New detectors with a different geometry, such as GridPix with its square pixels, will have to demonstrate the same. A partial overlap of the bands of adjacent instruments, such as 6--10~keV between the low- and medium-energy polarimeters of EXPO, or 20--30~keV between photoelectric and Compton stages, is therefore valuable beyond the gain in sensitivity, as it offers a cross-check of the systematic effects across different detectors and, in the second case, different techniques.

\paragraph{Spectroscopy and repointing.} Polarimeters are photon starved and their energy resolution is modest, so that fitting the polarisation of individual spectral components requires simultaneous, better-resolved spectra of the same source in the same time intervals; both EXPO, with its spectral imaging camera, and PHEMTO, with its Si and CdTe detectors, build this in. The time domain is the second dimension in which a dedicated mission can progress over IXPE, whose pointing programme is limited by its slow re-pointing.

\paragraph{Small and medium missions.} The soft band can be addressed by small missions such as GOSoX, since the Bragg technique profits from the high modulation, but not from a large area. The classical and hard bands, instead, need the effective area of M-class mirrors and the focal lengths that come with them. These two approaches are complementary: a mission of the EXPO type observing together with GOSoX and, perhaps, with a Compton-based mission, would span more than two decades in energy, and the open issues listed in Sect.~\ref{sec:tech}, such as the adsorption of DME, the Si$_3$N$_4$ protection layer of GridPix, the low-threshold active Compton polarimeter and the background rejection, are the ones on which the next years of laboratory work should concentrate.

\medskip
In summary, polarimetry below 2~keV and above 8~keV addresses questions that IXPE has raised but cannot answer: the thermal and QED physics of neutron stars, the geometry of the reflection in AGN, the transition between the thermal and non-thermal components of magnetars, and the non-thermal emission of SNRs and PWNe. The techniques to widen the band by more than a decade at both ends exist, or are close to being demonstrated, and several projects are putting them to use. The coming years, with the first flights of the soft-band polarimeters, the continued success of IXPE and the next opportunities for medium-class missions, will show how far the band of X-ray polarimetry can be widened.



\begin{thebibliography}{99}

\bibitem{novick1972}
R.~Novick, M.~C.~Weisskopf, R.~Berthelsdorf, R.~Linke and R.~S.~Wolff, \textit{Detection of X-ray polarization of the Crab Nebula}, Astrophys.\ J.\ \textbf{174}, L1 (1972).

\bibitem{weisskopf1978}
M.~C.~Weisskopf, E.~H.~Silver, H.~L.~Kestenbaum, K.~S.~Long and R.~Novick, \textit{A precision measurement of the X-ray polarization of the Crab Nebula without pulsar contamination}, Astrophys.\ J.\ \textbf{220}, L117 (1978).

\bibitem{weisskopf2022}
M.~C.~Weisskopf et al., \textit{The Imaging X-ray Polarimetry Explorer (IXPE): Pre-launch}, J.\ Astron.\ Telesc.\ Instrum.\ Syst.\ \textbf{8}, 026002 (2022).

\bibitem{soffitta2021instrument}
P.~Soffitta et al., \textit{The Instrument of the Imaging X-ray Polarimetry Explorer}, Astron.\ J.\ \textbf{162}, 208 (2021).

\bibitem{baldini2021}
L.~Baldini et al., \textit{Design, construction, and test of the Gas Pixel Detectors for the IXPE mission}, Astropart.\ Phys.\ \textbf{133}, 102628 (2021).

\bibitem{soffitta2024galaxies}
P.~Soffitta, E.~Costa, N.~De Angelis et al., \textit{Considerations on possible directions for a wide band polarimetry X-ray mission}, Galaxies \textbf{12}, 47 (2024).

\bibitem{soffitta2026lessons}
P.~Soffitta et al., \textit{Unlocking the future of X-ray polarimetry with IXPE: lessons learned and next steps}, arXiv:2601.00764 (2026).

\bibitem{bucciantini2023}
N.~Bucciantini et al., \textit{Simultaneous space and phase resolved X-ray polarimetry of the Crab pulsar and nebula}, Nat.\ Astron.\ \textbf{7}, 602 (2023).

\bibitem{vink2022}
J.~Vink et al., \textit{X-ray polarization detection of Cassiopeia A with IXPE}, Astrophys.\ J.\ \textbf{938}, 40 (2022).

\bibitem{taverna2022}
R.~Taverna et al., \textit{Polarized X-rays from a magnetar}, Science \textbf{378}, 646 (2022).

\bibitem{doroshenko2022}
V.~Doroshenko et al., \textit{Determination of X-ray pulsar geometry with IXPE polarimetry}, Nat.\ Astron.\ \textbf{6}, 1433 (2022).

\bibitem{liodakis2022}
I.~Liodakis et al., \textit{Polarized blazar X-rays imply particle acceleration in shocks}, Nature \textbf{611}, 677 (2022).

\bibitem{digesu2022}
L.~Di Gesu et al., \textit{The X-ray polarization view of Mrk 421 in an average flux state as observed by the Imaging X-ray Polarimetry Explorer}, Astrophys.\ J.\ Lett.\ \textbf{938}, L7 (2022).

\bibitem{gianolli2023}
V.~E.~Gianolli et al., \textit{Uncovering the geometry of the hot X-ray corona in the Seyfert galaxy NGC 4151 with IXPE}, Mon.\ Not.\ R.\ Astron.\ Soc.\ \textbf{523}, 4468 (2023).

\bibitem{ursini2023}
F.~Ursini et al., \textit{Mapping the circumnuclear regions of the Circinus galaxy with the Imaging X-ray Polarimetry Explorer}, Mon.\ Not.\ R.\ Astron.\ Soc.\ \textbf{519}, 50 (2023).

\bibitem{marin2023}
F.~Marin et al., \textit{X-ray polarization evidence for a 200-year-old flare of Sgr~A$^*$}, Nature \textbf{619}, 41 (2023).

\bibitem{pavlov2000}
G.~G.~Pavlov and V.~E.~Zavlin, \textit{Polarization of thermal X-rays from isolated neutron stars}, Astrophys.\ J.\ \textbf{529}, 1011 (2000).

\bibitem{zane2023}
S.~Zane et al., \textit{A strong X-ray polarization signal from the magnetar 1RXS J170849.0$-$400910}, Astrophys.\ J.\ Lett.\ \textbf{944}, L27 (2023).

\bibitem{turolla2023}
R.~Turolla et al., \textit{IXPE and XMM-Newton observations of the soft gamma repeater SGR 1806$-$20}, Astrophys.\ J.\ \textbf{954}, 88 (2023).

\bibitem{heyl2024}
J.~Heyl et al., \textit{The detection of polarized X-ray emission from the magnetar 1E 2259+586}, Mon.\ Not.\ R.\ Astron.\ Soc.\ \textbf{527}, 12219 (2024).

\bibitem{rigoselli2025}
M.~Rigoselli et al., \textit{IXPE detection of highly polarized X-rays from the magnetar 1E 1841$-$045}, Astrophys.\ J.\ Lett.\ \textbf{985}, L34 (2025).

\bibitem{stewart2025}
R.~Stewart et al., \textit{X-ray polarization of the magnetar 1E 1841$-$045}, Astrophys.\ J.\ Lett.\ \textbf{985}, L35 (2025).

\bibitem{stewart2026}
R.~E.~Stewart et al., \textit{Vacuum birefringence and the polarized X-ray emission from a radio magnetar}, Nature (2026), doi:10.1038/s41586-026-10859-z, arXiv:2509.19446.

\bibitem{taverna2026}
R.~Taverna et al., \textit{The long quest for vacuum birefringence in magnetars: 1E 1547.0$-$5408 and the elusive smoking gun}, Astrophys.\ J.\ \textbf{1002}, 102 (2026).

\bibitem{taverna2026ep}
R.~Taverna et al., \textit{IXPE observation of the new magnetar source EP J223759.5+531421}, arXiv:2609.25958 (2026).

\bibitem{davis2009}
S.~W.~Davis, O.~M.~Blaes, S.~Hirose and J.~H.~Krolik, \textit{The effects of magnetic fields and inhomogeneities on accretion disk spectra and polarization}, Astrophys.\ J.\ \textbf{703}, 569 (2009).

\bibitem{digesu2023}
L.~Di Gesu et al., \textit{Discovery of X-ray polarization angle rotation in the jet from blazar Mrk 421}, Nat.\ Astron.\ \textbf{7}, 1245 (2023).

\bibitem{zhang2013}
H.~Zhang and M.~B\"ottcher, \textit{X-ray and gamma-ray polarization in leptonic and hadronic jet models of blazars}, Astrophys.\ J.\ \textbf{774}, 18 (2013).

\bibitem{soffitta2026expo}
P.~Soffitta et al., \textit{EXPO: a quantum leap in fast, wide-band X-ray polarimetry for astrophysics}, Proc.\ SPIE (2026), arXiv:2608.20898.

\bibitem{krawczynski2011}
H.~Krawczynski et al., \textit{Scientific prospects for hard X-ray polarimetry}, Astropart.\ Phys.\ \textbf{34}, 550 (2011).

\bibitem{marin2024ngc1068}
F.~Marin et al., \textit{X-ray polarization measurement of the gold standard of radio-quiet active galactic nuclei: NGC 1068}, arXiv:2403.02061 (2024).

\bibitem{goosmann2011}
R.~W.~Goosmann and G.~Matt, \textit{Spotting the misaligned outflows in NGC 1068 using X-ray polarimetry}, Mon.\ Not.\ R.\ Astron.\ Soc.\ \textbf{415}, 3119 (2011).

\bibitem{grefenstette2015}
B.~W.~Grefenstette et al., \textit{Locating the most energetic electrons in Cassiopeia A}, Astrophys.\ J.\ \textbf{802}, 15 (2015).

\bibitem{gotz2006}
D.~G\"otz, S.~Mereghetti, A.~Tiengo and P.~Esposito, \textit{Magnetars as persistent hard X-ray sources: INTEGRAL discovery of a hard tail in SGR 1900+14}, Astron.\ Astrophys.\ \textbf{449}, L31 (2006).

\bibitem{makishima2021}
K.~Makishima, T.~Enoto, H.~Yoneda and H.~Odaka, \textit{A NuSTAR confirmation of the 36 ks hard X-ray pulse-phase modulation in the magnetar 1E 1547.0$-$5408}, Mon.\ Not.\ R.\ Astron.\ Soc.\ \textbf{502}, 2266 (2021).

\bibitem{meszaros1988}
P.~M\'esz\'aros, R.~Novick, A.~Szentgyorgyi, G.~A.~Chanan and M.~C.~Weisskopf, \textit{Astrophysical implications and observational prospects of X-ray polarimetry}, Astrophys.\ J.\ \textbf{324}, 1056 (1988).

\bibitem{caiazzo2021a}
I.~Caiazzo and J.~Heyl, \textit{Polarization of accreting X-ray pulsars -- I. A new model}, Mon.\ Not.\ R.\ Astron.\ Soc.\ \textbf{501}, 109 (2021).

\bibitem{caiazzo2021b}
I.~Caiazzo and J.~Heyl, \textit{Polarization of accreting X-ray pulsars -- II. Hercules X-1}, Mon.\ Not.\ R.\ Astron.\ Soc.\ \textbf{501}, 129 (2021).

\bibitem{abarr2021}
Q.~Abarr et al., \textit{XL-Calibur -- a second-generation balloon-borne hard X-ray polarimetry mission}, Astropart.\ Phys.\ \textbf{126}, 102529 (2021).

\bibitem{awaki2025crab}
H.~Awaki et al., \textit{XL-Calibur measurements of polarised hard X-ray emission from the Crab}, Mon.\ Not.\ R.\ Astron.\ Soc.\ Lett.\ (2025), doi:10.1093/mnrasl/slaf026, arXiv:2503.14307.

\bibitem{awaki2025cyg}
H.~Awaki et al., \textit{XL-Calibur polarimetry of Cyg X-1 further constrains the origin of its hard-state X-ray emission}, arXiv:2507.23126 (2025).

\bibitem{laurent2011}
P.~Laurent, J.~Rodriguez, J.~Wilms, M.~Cadolle Bel, K.~Pottschmidt and V.~Grinberg, \textit{Polarized gamma-ray emission from the Galactic black hole Cygnus X-1}, Science \textbf{332}, 438 (2011).

\bibitem{krawczynski2022}
H.~Krawczynski et al., \textit{Polarized x-rays constrain the disk-jet geometry in the black hole x-ray binary Cygnus X-1}, Science \textbf{378}, 650 (2022).

\bibitem{weisskopf2010}
M.~C.~Weisskopf, R.~F.~Elsner and S.~L.~O'Dell, \textit{On understanding the figures of merit for detection and measurement of X-ray polarization}, Proc.\ SPIE \textbf{7732} (2010), arXiv:1006.3711.

\bibitem{marshall2017}
H.~L.~Marshall et al., \textit{The rocket experiment demonstration of a soft X-ray polarimeter (REDSoX Polarimeter)}, Proc.\ SPIE \textbf{10397}, 103970K (2017).

\bibitem{heine2026}
S.~N.~T.~Heine et al., \textit{Design and testing progress towards the first flight of the rocket experiment demonstration of a Soft X-ray Polarimeter (REDSoX)}, arXiv:2607.27639 (2026).

\bibitem{marshall2026}
H.~L.~Marshall et al., \textit{A new NASA Pioneer: the Globe Orbiting Soft X-ray Polarimeter (GOSoX)}, arXiv:2608.02891 (2026).

\bibitem{pacciani2003}
L.~Pacciani et al., \textit{Sensitivity of a photoelectric X-ray polarimeter for astronomy: the impact of the gas mixture and pressure}, Proc.\ SPIE \textbf{4843}, 394 (2003).

\bibitem{costa2001}
E.~Costa et al., \textit{An efficient photoelectric X-ray polarimeter for the study of black holes and neutron stars}, Nature \textbf{411}, 662 (2001).

\bibitem{bellazzini2007}
R.~Bellazzini et al., \textit{A sealed Gas Pixel Detector for X-ray astronomy}, Nucl.\ Instrum.\ Methods Phys.\ Res.\ A \textbf{579}, 853 (2007).

\bibitem{minuti2023}
M.~Minuti et al., \textit{XPOL-III: a new-generation VLSI CMOS ASIC for high-throughput X-ray polarimetry}, Nucl.\ Instrum.\ Methods Phys.\ Res.\ A \textbf{1046}, 167674 (2023).

\bibitem{poikela2014}
T.~Poikela et al., \textit{Timepix3: a 65K channel hybrid pixel readout chip with simultaneous ToA/ToT and sparse readout}, J.\ Instrum.\ \textbf{9}, C05013 (2014).

\bibitem{llopart2022}
X.~Llopart et al., \textit{Timepix4, a large area pixel detector readout chip which can be tiled on 4 sides providing sub-200 ps timestamp binning}, J.\ Instrum.\ \textbf{17}, C01044 (2022).

\bibitem{vandergraaf2007}
H.~van der Graaf, \textit{GridPix: an integrated readout system for gaseous detectors with a pixel chip as anode}, Nucl.\ Instrum.\ Methods Phys.\ Res.\ A \textbf{580}, 1023 (2007).

\bibitem{ligtenberg2018}
C.~Ligtenberg et al., \textit{Performance of a GridPix detector based on the Timepix3 chip}, Nucl.\ Instrum.\ Methods Phys.\ Res.\ A \textbf{908}, 18 (2018).

\bibitem{black2007}
J.~K.~Black et al., \textit{X-ray polarimetry with a micropattern TPC}, Nucl.\ Instrum.\ Methods Phys.\ Res.\ A \textbf{581}, 755 (2007).

\bibitem{dimarco2023}
A.~Di Marco et al., \textit{Handling the background in IXPE polarimetric data}, Astron.\ J.\ \textbf{165}, 143 (2023).

\bibitem{hill2012}
J.~E.~Hill et al., \textit{The design and qualification of the GEMS X-ray polarimeters}, Proc.\ SPIE \textbf{8443} (2012).

\bibitem{fabiani2012}
S.~Fabiani et al., \textit{Performance of an Ar-DME imaging photoelectric polarimeter}, Proc.\ SPIE \textbf{8443} (2012).

\bibitem{soffitta2010}
P.~Soffitta et al., \textit{A set of X-ray polarimeters for the New Hard X-ray Imaging and Polarimetric Mission}, Proc.\ SPIE \textbf{7732}, 77321A (2010).

\bibitem{tagliaferri2012}
G.~Tagliaferri et al., \textit{The NHXM observatory}, Exp.\ Astron.\ \textbf{34}, 463 (2012).

\bibitem{takeda2020}
T.~Takeda et al., \textit{Development and performance verification of a TPC polarimeter for high energy X-rays}, Proc.\ SPIE \textbf{11444}, 114445Y (2020).

\bibitem{abarr2020}
Q.~Abarr et al., \textit{Observations of a GX 301-2 apastron flare with the X-Calibur hard X-ray polarimeter supported by NICER, the Swift XRT and BAT, and Fermi GBM}, Astrophys.\ J.\ \textbf{891}, 70 (2020).

\bibitem{hayashida2016}
K.~Hayashida et al., \textit{Hard X-ray imaging polarimeter for PolariS}, Proc.\ SPIE \textbf{9905}, 99051A (2016).

\bibitem{krawczynski2016}
H.~S.~Krawczynski et al., \textit{X-ray polarimetry with the Polarization Spectroscopic Telescope Array (PolSTAR)}, Astropart.\ Phys.\ \textbf{75}, 8 (2016).

\bibitem{delmonte2023}
E.~Del Monte, S.~Fabiani and M.~Pearce, \textit{Compton polarimetry}, in \textit{Handbook of X-ray and Gamma-ray Astrophysics}, eds.\ C.~Bambi and A.~Santangelo (Springer, 2023), p.~127.

\bibitem{costa2024}
E.~Costa, \textit{Scattering polarimetry in the hard X-ray range}, Instruments \textbf{8}, 20 (2024).

\bibitem{fabiani2013}
S.~Fabiani et al., \textit{Characterization of scatterers for an active focal plane Compton polarimeter}, Astropart.\ Phys.\ \textbf{44}, 91 (2013).

\bibitem{iyer2023}
N.~K.~Iyer et al., \textit{The design and performance of the XL-Calibur anticoincidence shield}, Nucl.\ Instrum.\ Methods Phys.\ Res.\ A \textbf{1048}, 167975 (2023).

\bibitem{muleri2024}
F.~Muleri et al., \textit{In search of the third dimension in Compton X-ray polarimeters}, Proc.\ SPIE \textbf{13093} (2024).

\bibitem{laurent2021}
P.~Laurent et al., \textit{PHEMTO: the polarimetric high energy modular telescope observatory}, Exp.\ Astron.\ \textbf{51}, 1143 (2021).

\bibitem{esa2026}
ESA, \textit{Seven-spacecraft observatory to study plasma universe}, ESA Science news, 19 November 2026.

\bibitem{jahoda2019}
K.~Jahoda et al., \textit{The X-ray Polarization Probe mission concept}, arXiv:1907.10190 (2019).

\bibitem{soffitta2021ngxp}
P.~Soffitta et al., \textit{A polarized view of the hot and violent universe}, Exp.\ Astron.\ \textbf{51}, 1109 (2021).

\bibitem{muleri2026}
F.~Muleri et al., \textit{Instruments for focal plane X-ray polarimetry in the next decade}, Particles (2026), arXiv:2606.20337.

\bibitem{zhang2019}
S.~N.~Zhang et al., \textit{eXTP -- enhanced X-ray Timing and Polarimetry mission}, Sci.\ China Phys.\ Mech.\ Astron.\ \textbf{62}, 29502 (2019).

\bibitem{zhou2025}
P.~Zhou et al., \textit{Observatory science with eXTP}, arXiv:2506.08367 (2025).

\bibitem{yin2022}
Q.-Q.~Yin, Z.-W.~Li and C.-Z.~Liu, \textit{The wide band X-ray polarization telescope (WXPT)}, 44th COSPAR Scientific Assembly, Vol.~44, p.~1874 (2022), abstract E1.12-0031-22, ADS 2022cosp...44.1874Y.

\bibitem{cai2026}
Z.~Cai et al., \textit{Deposition and performance of depth-graded W/Si multilayers on silicon pore optics for the WXPT mission}, Exp.\ Astron.\ \textbf{62}, 2 (2026), doi:10.1007/s10686-026-10065-9.

\bibitem{lundman2018}
C.~Lundman, I.~Vurm and A.~M.~Beloborodov, \textit{Polarization of gamma-ray bursts in the dissipative photosphere model}, Astrophys.\ J.\ \textbf{856}, 145 (2018).

\bibitem{gill2020}
R.~Gill, J.~Granot and P.~Kumar, \textit{Linear polarization in gamma-ray burst prompt emission}, Mon.\ Not.\ R.\ Astron.\ Soc.\ \textbf{491}, 3343 (2020).

\bibitem{lan2021}
M.-X.~Lan, H.-B.~Wang, S.~Xu, S.~Liu and X.-F.~Wu, \textit{Polarization predictions in the GRB prompt phase with the internal shock model}, Astrophys.\ J.\ \textbf{909}, 184 (2021).

\bibitem{kole2026}
M.~Kole, N.~De Angelis, J.~He, H.~Liu, J.~Sun, F.~Xie and J.~Zaid, \textit{Design and scientific prospects of the POLAR-2 mission}, Innov.\ Space Res.\ Technol.\ \textbf{1}, 2 (2026), arXiv:2510.02016.

\bibitem{deangelis2023}
N.~De Angelis, \textit{Development of the next generation space-based Compton polarimeter and energy resolved polarization analysis of gamma-ray bursts prompt emission}, PhD thesis, University of Geneva (2023), doi:10.13097/archive-ouverte/unige:173869.

\bibitem{gros2018}
P.~Gros et al., \textit{Performance measurement of HARPO: a time projection chamber as a gamma-ray telescope and polarimeter}, Astropart.\ Phys.\ (2018), doi:10.1016/j.astropartphys.2017.10.008, arXiv:1706.06483.

\bibitem{nakamura2024}
Y.~Nakamura et al., \textit{New high-precision measurement system for electron-positron pairs from sub-GeV/GeV gamma-rays in the emulsion telescope}, arXiv:2406.18335 (2024).

\end{thebibliography}
\end{document}